\documentclass[twocolumn,aps,prl,showpacs,preprintnumbers,amsmath,amssymb, superscriptaddress]{revtex4}
\usepackage{graphicx}% Include figure files
\usepackage{dcolumn}% Align table columns on decimal point
\usepackage{bm}% bold math

\begin{document}

%\preprint{}

\title{An Apparent Dissociation Transition in Anharmonically Bound 1D Systems} 


\author{D. J. Priour, Jr}
\affiliation{Department of Physics \& Astronomy, Youngstown State University, Youngstown, OH 44555, USA}
\author{Christopher Watenpool}
\affiliation{Department of Electrical Engineering, Youngstown State University, Youngstown, OH 44555, USA}
%\thanks{}


\date{\today}

\begin{abstract}
For diatomic molecules and chains bound anharmonically by interactions such as the 
Lennard Jones and Morse potentials, we obtain analytical expressions for thermodynamic 
observables including the mean bond length, thermally averaged internal energy, 
and the coefficient of thermal expansion.  These results are valid across the 
shift from condensed to gas-like phases, a dissociation transition 
marked by a crossover with no singularities in thermodynamic variables 
for finite pressures, though singular behavior appears in the low pressure limit.
In the regime where the thermal energy $k_{\mathrm{B}} T$ is much smaller than the 
dissociation energy $D$, the mean interatomic separation  
scales as $\langle l \rangle = R_{e} + {\mathcal B} (P R_{e}/k_{\mathrm{B}} T)^{-2} e^{-D/k_{\mathrm{B}} T} \left( D/k_{\mathrm{B}} T \right )^{1/2}$
for both the Morse and Lennard Jones potentials
where $p$ is a pressure term, $R_{e}$ is the $T = 0$ bond length, and ${\mathcal B}$ is a constant specific to the potential.
\end{abstract}
%\keywords{}

\maketitle

%\section{}
%\subsection{}
%\subsubsection{}

Potentials used to describe interactions among atoms such as the Lennard Jones potential~\cite{Lennard}
(e.g. Van der Waals couplings among noble gas atoms such as Argon) and the 
Morse potential~\cite{Morse} (for bonding with significant covalent character) contain attractive 
and repulsive terms.
Whereas the attractive component decays at large distance, the repulsive piece decays even more rapidly 
in the large separation regime while rising sharply when atoms are in proximity where Pauli Exclusion effects play 
a role as atomic cores begin to overlap.  With the combination of attractive and repulsive terms, there is a 
potential minimum at the equilibrium separation $R_{e}$ with harmonic (parabolic) dependence in the vicinity of 
$R_{e}$ and increasingly asymmetric and anharmonic character for significant deviations from $R_{e}$ caused, e.g., by 
thermal fluctuations in the high temperature regime.  Whether due to Van der Waals coupling as in the Lennard Jones model or 
the sharing of charge represented by the Morse potential, entropic effects drive dissociation despite the attractive component,
except at $T = 0$ where thermal fluctuations are absent.  Nevertheless, confinement of the system volume 
for finite temperatures is realized by taking into consideration a pressure term (incorporated in the 1D 
context as a finite cost per length of elongating the system, such as a chain of identical atoms), which precludes 
indefinite expansion of the system.

In this work, we operate in terms of three distinct energy 
scales relevant to both the Morse and Lennard Jones potentials; the thermal energy $k_{\mathrm{B}} T$, the dissociation energy $D$ needed to completely separate an 
interacting pair of atoms, and the pressure confinement scale $p R_{e}$.  For high $T$, where $k_{\mathrm{B}} T \gg D$ the competition among
$p R_{e}$ and $k_{\mathrm{B}} T$ dominates, and the state is well approximated as an ideal gas.  On the other hand,
for low $T$, where $k_{\mathrm{B}} T \ll D$, there is a more subtle interplay among $p$, $T$, and $D$ with a crossover 
with decreasing $p$ from a condensed state 
($ \langle l \rangle \approx R_{e}$) to a diluted gas-like state ($\langle l \rangle \gg R_{e}$).
For a more detailed description of dissociation, we  obtain analytical expressions for thermodynamic variables of interest valid 
across the shift from the condensed to gas-like states in the low to intermediate temperature regimes.
With a unified treatment applied to both the Lennard Jones and Morse cases, we find strong qualitative similarities in 
molecular dissociation phenomena despite distinct bonding physics driving the potentials.

With interactions only among nearest neighbors, the system energy is $E = \sum_{i} P_{i}^{2}/2M + \sum_{i} V (R_{i+1} - R_{i}) + pL$ 
with the sum over the $N$ members of the chain, $L$ being the total system size.  Since atomic momenta and 
spatial coordinates are not coupled, apart from a contribution $N k_{\mathrm{B}} T/2$ to the internal energy (and $N k_{\mathrm{B}}/2$ 
to the specific heat), one need only consider site positions in sampling system configurations. We obtain
thermodynamic variables of interest from the partition function $Z$; though in general the rapid scaling of 
configuration space with the number of system components precludes an analytical calculation of $Z$, we decouple the 
calculation of $Z$ by describing the system in terms of the separation among adjacent sites $\Delta_{i} \equiv R_{i+1} - R_{i}$ 
instead of the absolute atomic coordinates $R_{i}$~\cite{MacDonald1,MacDonald2}.
Noting that $L = R_{N} - R_{1} = \Delta_{N-1} + \Delta_{N-2} + \ldots + \Delta_{1}$, the partition function may be 
factorized as $Z = \hat{Z}^{N-1}$ where $\hat{Z} = \int_{0}^{\infty} \exp ( -\beta [ V(\Delta) + p \Delta ] ) d \Delta$. In this manner, 
the calculation for a chain of arbitrary length is reduced to the case of a single atomic pair separated by a distance $\Delta$ (subsequently
labeled as $R$), subject to an asymptotically linearly diverging effective potential $V(R) + pR$.

It is convenient to operate in terms of dimensionless energy variables $\varepsilon \equiv \beta D$ and $\eta \equiv p R_{e}$ as well 
as the ratio $\gamma \equiv p R_{e}/D = \eta/\varepsilon$ characterizing the strength of the pressure energy scale in relation to
the dissociation energy.  Salient 
thermodynamic variables may then be calculated by differentiating with respect to $\varepsilon$ and $\eta$, such as 
$\langle E \rangle = -\partial/\partial \beta \ln Z = -( \varepsilon Z_{\varepsilon} + \eta Z_{\eta} ) \beta^{-1} Z^{-1}$ 
for the internal energy where $Z_{\varepsilon} \equiv \partial Z/\partial \varepsilon$ and $Z_{\eta} \equiv \partial Z/\partial \eta$.  
Though having different functional forms, $V_{\mathrm{LJ}} = B R^{-12} - A R^{-6}$ (Lennard Jones) and $V_{\mathrm{M}} = -D + D \{-1 + \exp[-a (R - R_{e})] \}^{2}$
(Morse), $V_{\mathrm{LJ}}$ and $V_{\mathrm{M}}$ are amenable to the same analysis with qualitatively similar results in both cases.
   
When expressed in terms of $D$ and the reduced 
separation $r \equiv R/R_{e}$, $V_{\mathrm{LJ}}$ and $V_{\mathrm{M}}$ are both of the form
$V(r) = -D + D[ \chi(r) - 1]^{2}$ where $\chi_{\mathrm{LJ}} = r^{-6}$ and $\chi_{\mathrm{M}} = \exp[- {\mathcal A} (r - 1)]$.
In terms of specific parameters, $D = -A^{2}/4B$ and $R_{e} = (2B/A)^{1/6}$ for the Lennard Jones case while 
${\mathcal A} = a R_{e}$ (${\mathcal A}  = 2.5$ for results exhibited here) for 
the Morse potential.  In terms of $\chi(r)$, the partition function for both 
$V_{\mathrm{LJ}}$ and $V_{\mathrm{M}}$ is $Z = R_{e} \exp{\varepsilon} \int_{0}^{\infty} \exp[ \chi(r) -1]^{2} \exp[-\eta r] dr$.
 
Although one could in principle expand $\chi(r) - 1$ about $r = 1$, where the potential basin is locally parabolic,   
the significant departure from harmonic character with the shift from the condensed phase ($r \approx 1$) to the gas-like phase 
($r \gg 1$) hampers a perturbative analysis of this kind.
For a treatment which offers good convergence with only a handful of terms, and which provides a 
good description of the condensed and dissociated states alike, we 
consider the substitution $u = \chi(r)$, obtaining 
\begin{equation}
Z = -R_{e} e^{\varepsilon} \int_{0}^{\infty} e^{-\varepsilon (u - 1)^{2}} e^{-\eta \chi^{-1}(u)} \frac{1}{\frac{d}{du} \chi^{-1}(u)} du
\end{equation}  
Integrating by parts, neglecting boundary terms, and using $v = u - 1$ leads to 
\begin{equation}
Z = \frac{2 R_{e} \varepsilon}{\eta} e^{\varepsilon} \int_{-1}^{\infty} e^{-\varepsilon v^{2}} e^{-\eta \chi^{-1}(v+1)} v dv
\end{equation}
which may be evaluated as a series of Gaussian integrals by Taylor expanding $\exp [-\eta \chi^{-1}(v+1)]$ about $v = 0$
with 
\begin{equation}
e^{-\eta \chi^{-1}(v+1)} = e^{-\eta} \left[ 1 - \eta d_{1} v + \frac{1}{2!} (\eta^{2} d_{1}^{2} - \eta d_{2})v^{2} + \ldots \right ]
\end{equation}
where the $d_{i}$ are $i$th derivatives of $\chi^{-1}(v+1)$ evaluated at $v = 0$. For $V_{\mathrm{LJ}}$, $d_{1} = -1/6$, $d_{2} = 7/36$, and 
$d_{3} = -91/216$ while for $V_{\mathrm{M}}$, $d_{1} = -1/{\mathcal A}$, $d_{2} = 1/{\mathcal A}$, and $d_{3} = -2/{\mathcal A}$.

For the sake of a quantitatively accurate description of the dissociation transition, it is important not to neglect the 
finiteness of the lower integration limit in evaluating the Gaussian integrals, which must be taken into consideration to
account for dissociation.  In the regime of interest, $Z$ is well represented by
\begin{eqnarray}
Z = e^{\varepsilon - \eta} \bigg[ e^{-\varepsilon} (\eta^{-1} + d_{1} ) - \\ 
\nonumber
\frac{\sqrt{\pi}}{4} \varepsilon^{-3/2} (4 \varepsilon d_{1} + d_{1}^{3} \eta^{2} - 
3 \eta d_{1} d_{2} + d_{3} ) \bigg]
\end{eqnarray}
where we have used $\int_{-1}^{\infty} \exp( -\varepsilon v^{2})v dv = \exp -\varepsilon/(2 \varepsilon)$, 
 $\int_{-1}^{\infty} \exp( -\varepsilon v^{2})v^{2} dv \approx (\sqrt{\pi}/2) \varepsilon^{-3/2} - \exp -\varepsilon/(2 \varepsilon)$ 
(i.e. retaining the leading term in the 
asymptotic series); we have neglected $\int_{-1}^{\infty} \exp( -\varepsilon v^{2})v^{3} dv$, and truncated 
$\int_{-1}^{\infty} \exp( -\varepsilon v^{2})v^{4} dv$ at $(3 \sqrt{\pi}/4) \varepsilon^{-5/2}$.

For a description valid for both the condensed and gas-like phases,  we obtain from $Z$ rational expressions for 
observables of interest where typically leading and next to leading order 
terms in $\eta$ and $\varepsilon$ are retained in the numerator and the denominator; 
we calculate and discuss in turn the thermally averaged bond length $\langle l \rangle$, the internal energy $\langle E \rangle$, 
the specific heat $c_{p}$, and the thermal expansion coefficient $\alpha$.

With the mean interatomic separation being $-R_{e} \partial \ln Z/\partial \eta$, the normalized thermally averaged bond length is
well approximated with
\begin{equation}
\frac{\langle l \rangle}{R_{e}} = 1 + \eta^{-1} \left [ \frac{e^{-\varepsilon} - \frac{3}{4} \sqrt{\pi} d_{1} d_{2} \varepsilon^{-3/2} \eta^{2} }
{e^{-\varepsilon} -\frac{\sqrt{\pi}}{4} \varepsilon^{-3/2} \eta \left( 4 d_{1} \varepsilon +   d_{3} \right )} \right ]
\label{eq:Eq5}
\end{equation}
readily inverted via the quadratic formula to obtain $\eta$ in terms of $\langle l \rangle$.  Asymptotically, for 
$\eta \ll \varepsilon$, the residual component of the mean separation (small in the condensed phase but diverging with the dissociation transition)
may be represented by the simpler expression 
$\tilde{\Delta}_{l} \equiv (\langle l \rangle/R_{e} - 1) = \varepsilon^{1/2} \exp -\varepsilon /(d_{1} \sqrt{\pi} \eta^{2})$.

To more conveniently visualize $\langle l \rangle$, (as well as $\langle E \rangle$, $\alpha$, and $c_{p}$), we choose $\varepsilon$ for the abscissa
where only $T$ is varied
while holding $p$ fixed for a given curve; $\varepsilon \gg 1$  and $\varepsilon \ll 1$ correspond to the low and high $T$ regimes respectively.
Whereas $\gamma = p R_{e}/D$ remains constant, $\eta = \beta p R_{e} = \gamma \varepsilon$ varies with temperature.  
         
\begin{figure}
\includegraphics[width=.5\textwidth]{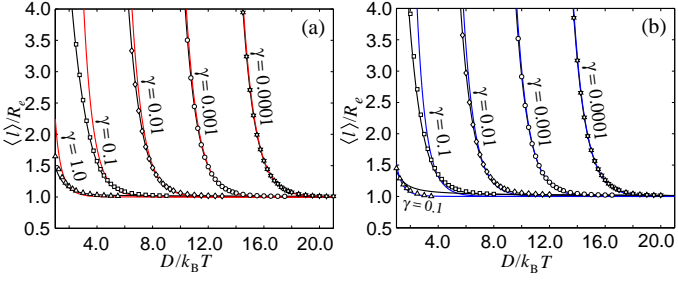}
\caption{\label{fig:Fig1} (Color online) Normalized mean atomic separations for the Lennard Jones (left panel) and Morse potential (right panel)
for various $\gamma$ values; 
solid traces are analytical approximations while open symbols are numerical data.  Black curves correspond to  the $\tilde{\Delta}_{l}+1$ 
relationship, while colored traces represent the rational $\langle l \rangle$ expression.}
\end{figure}

Figure~\ref{fig:Fig1} displays results corresponding to $V_{\mathrm{LJ}}$ in panel (a) and $V_{\mathrm{M}}$ in panel (b)  
with open symbols indicating numerical results and solid curves representing the foregoing approximations to $\langle l \rangle$.
The black traces, calculated with the rational expression in Eq.~\ref{eq:Eq5}, are in good quantitative agreement with numerical data even for $\gamma = 1.0$. 
On the other hand, the $\tilde{\Delta}_{l} + 1$ approximation is represented by 
lighter (red/blue for  $V_{\mathrm{LJ}}$/$V_{\mathrm{M}}$) curves,
and accurately indicates the location where $\langle l \rangle$ begins to diverge significantly from 
$R_{e}$.  Moreover, despite the simple structure, good agreement with exact numerical results is evident for $\gamma < 0.01$.  

To estimate $\gamma$ for physically realistic systems, we consider a chain of noble gas atoms (e.g. Argon) in the context of the 
Lennard Jones Model or a covalently bonded pair of atoms  
(e.g. H$_{2}$) in the Morse potential framework in a 1D conduit. With typical atomic radii on the order of an \AA ngstrom,  
we assume a cross sectional area on the order of $(1~\textrm{\AA})^{2} = 10^{-20}~\textrm{m}^{2}$.  Using $1.01 \times 10^{5}~\textrm{Pa}$ for standard atmospheric 
pressure, find $p \sim 10^{-15}~J/\textrm{m}$.  In the case of Argon, one has $D = 0.011~\textrm{eV}$ and $R_{e} = 3.8~\textrm{\AA}$~\cite{Rahman,Barker,Rowley,John}  
and $\gamma = 2.2 \times 10^{-4}$.  On the other hand, for H$_{2}$, $D = 4.52~\textrm{eV}$~\cite{Blanksby} 
and we take $R_{e}$ to be the measured bond length of $0.74~\textrm{\AA}$~\cite{Gray}; one obtains 
$\gamma = 1.0 \times 10^{-7}$, with both $\gamma$ values deep in the $\gamma \ll 1$ range.  

The internal energy, $\langle E \rangle$, is well approximated as
\begin{equation}
\frac{\langle E \rangle}{k_{\mathrm{B}} T} = \eta - \varepsilon  + \frac{e^{-\varepsilon} \eta^{-1} (1 + \varepsilon ) - \varepsilon^{-3/2}
\frac{\sqrt{\pi}}{8} (4 d_{1} \varepsilon + 3d_{3} )}
{e^{-\varepsilon}\eta^{-1} - \varepsilon^{-3/2} \frac{\sqrt{\pi}}{4}(4 d_{1} \varepsilon +  d_{3}) }
\end{equation}

\begin{figure}
\includegraphics[width=.5\textwidth]{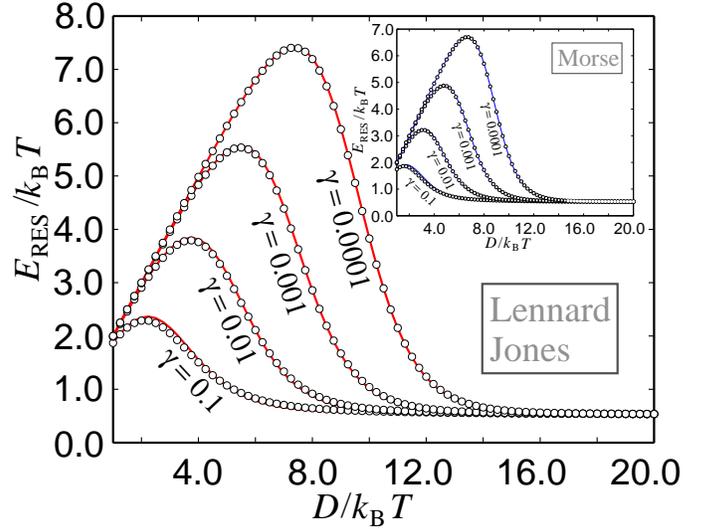}
\caption{\label{fig:Fig2} (Color online) Residual internal energy results for various $\gamma$ values with the main graph showing Lennard Jones results and the inset 
displaying Morse results.  Solid traces represent analytical results, while open symbols indicate numerical data.}
\end{figure}

To strip away trivial dependencies, Figure~\ref{fig:Fig2} shows the residual internal energies, $E_{\mathrm{Res}} = \langle E \rangle + D - p R_{e}$, 
with Lennard Jones results in the main panel and Morse potential results in the 
inset for a variety of $\gamma$ values.  Solid curves obtained from the rational expression closely coincide with the open symbols representing numerical data.
Though $E_{\mathrm{Res}}$ tends to $k_{\mathrm{B}} T/2$ for sufficiently large $\varepsilon$, the expected dependence where 
$\langle l \rangle \approx R_{e}$ where anharmonicities are negligible, the residual energy component is non-monotonic.  One sees from the 
main graph and the inset graph that the breadth and height of the peak separating the low and high $T$ regimes scales asymptotically as $\log_{10}(1/\gamma)$  

The specific heat at constant pressure is $c_{p} = \partial \langle E \rangle/\partial T$, which is
\begin{equation}
\frac{c_{p}}{k_{\mathrm{B}}} = \frac{\frac{1}{2} \varepsilon^{2} \left( \varepsilon + \frac{3 d_{3}}{2d_{1}} \right) 
- \Upsilon \varepsilon^{1/2} \left( 1 + \varepsilon + \frac{d3}{4d_{1}} \right )  + \Upsilon^{2}}
{\varepsilon^{-2}\left( \varepsilon + \frac{d_{3}}{2d_{1}} \right ) 
 - 2 \Upsilon \varepsilon^{-3/2} \left( \varepsilon + \frac{d3}{4d_{1}} \right )  + \Upsilon^{2}  }
\label{eq:Eq7}
\end{equation}
where $\Upsilon =  \exp{-\varepsilon}/( \sqrt{\pi} \eta d_{1} )$.

\begin{figure}
\includegraphics[width=.5\textwidth]{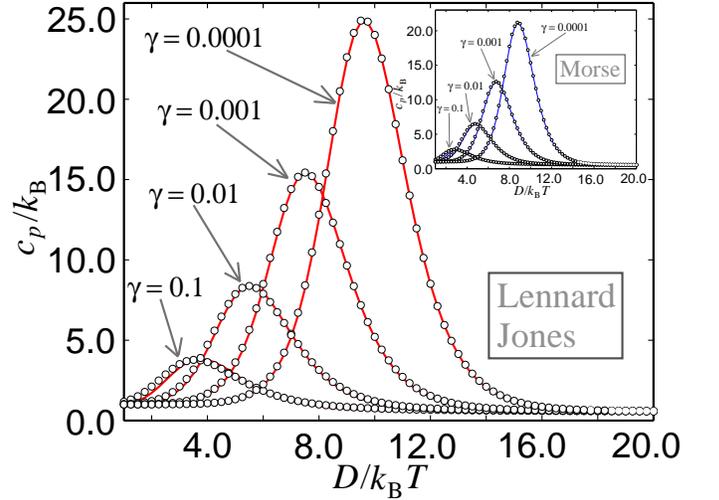}
\caption{\label{fig:Fig3} (Color online) Specific heat curves for assorted $\gamma$ values with Lennard Jones results in the main graph and Morse results in the    
inset.  Analytical results are shown as solid traces, while open symbols represent numerical data.}
\end{figure}

Specific heat results are displayed in the main graph (for $V_{\mathrm{LJ}}$) and the inset (for $V_{\mathrm{M}}$) of 
Fig.~\ref{fig:Fig3} for a range of $\gamma$ values; as in the case of the internal energy, there is good agreement among the analytical (solid curve) and 
the numerical (open symbols) results.    

One sees from the analytical $c_{p}$ expression in Eq.~\ref{eq:Eq7} and the curves in Fig.~\ref{fig:Fig3} that the specific heat tends to $k_{\mathrm{B}}$ for 
sufficiently low $D$ (i.e. higher $T$) and flattens to  $k_{\mathrm{B}}/2$ for $D \gg 1$ (low $T$).  Whereas the latter is a hallmark of the 
condensed phase, the former is expected for a dissociated system; the transition between the two asymptotically flat regions is non-monotonic, with a peak separating the 
$k_{\mathrm{B}}$ and $k_{\mathrm{B}}/2$ regimes.  The region where $c_{p}$ rises to a peak, representing the dissociation transition, becomes taller and  
narrower with decreasing $\gamma$.

The coefficient of thermal expansion $\alpha = \langle l \rangle^{-1} \partial \langle l \rangle/\partial T$ is well represented by
\begin{equation}
\frac{\alpha}{k_{\mathrm{B}}}  = \frac{\frac{3}{4} \eta d_{2} \varepsilon^{-1} -\Upsilon \varepsilon^{1/2} \left( \varepsilon + \frac{3}{2} 
+ \frac{d_{3}}{4 d_{1}} - \frac{3 \eta d_{2}}{2} \right )  + \Upsilon^{2}}{\eta -\Upsilon \varepsilon^{-1/2} \left( \varepsilon +2 \eta \varepsilon + \frac{d_{3}}{4 d_{1}}
\right )  + \Upsilon^{2}} 
\end{equation}

Juxtaposed analytical (solid traces) and numerical results (open symbols) are shown in Figure 4 for various $\gamma$ values with 
$D \alpha/k_{\mathrm{B}}$ on the vertical axis.   As in the case of the specific heat, the $\alpha$ curves 
are non-monotonic, with peak heights increasing with decreasing $\gamma$.  Choosing $D \alpha/k_{\mathrm{B}}$ for the ordinate is in 
part to show the slow convergence to the low $T$ value of $3 d_{2}/4$ in the  $\gamma \ll 1$ limit.  All of the cases shown correspond to 
experimentally realistic $\gamma$ values, and in all but one of the curves shown, $\alpha$ is appreciably different from the limiting value even 
for $\varepsilon$ as high as 10.
\begin{figure}
\includegraphics[width=.5\textwidth]{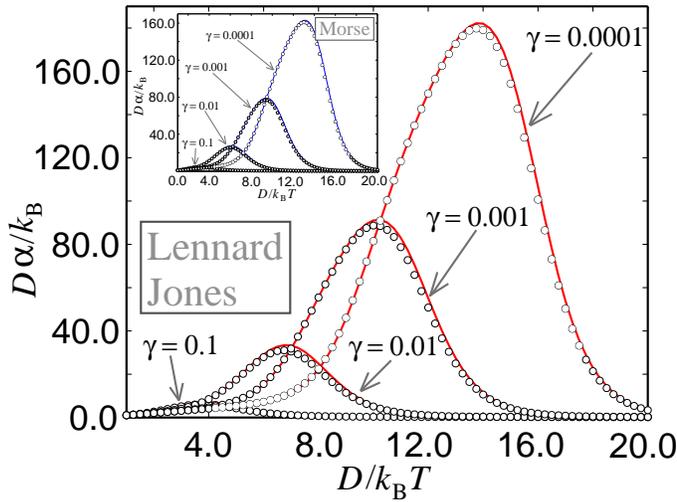}
\caption{\label{fig:Fig4} (Color online) Thermal expansion coefficient results for a range of $\gamma$ values with Lennard Jones results plotted in the main 
graph and corresponding results for the Morse potential in the inset.  Solid curves represent analytical results, while open symbols indicate numerical data.}
\end{figure}

\begin{figure}
\includegraphics[width=.5\textwidth]{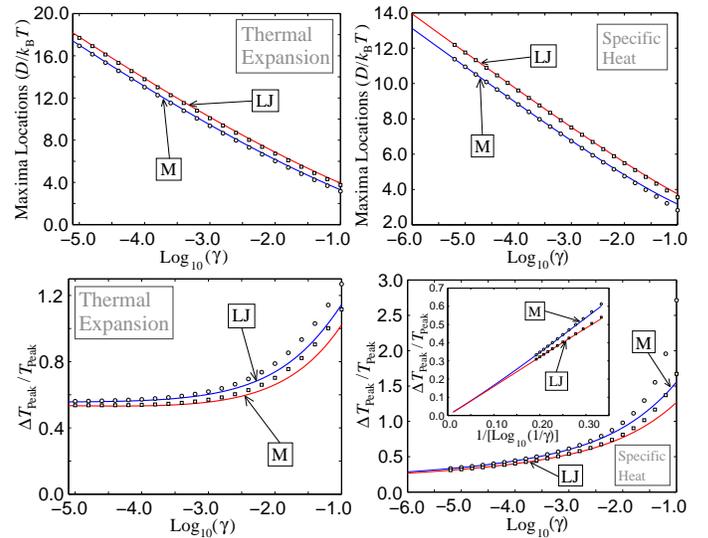}
\caption{\label{fig:Fig5} (Color online) Peak locations for the specific heat (upper right panel) and thermal expansion coefficient (upper left panel).
Corresponding relative full width half maxima (i.e. $\Delta T_{\mathrm{peak}}/T_{\mathrm{peak}}$) appear in the lower right and lower left panels for the 
specific heat and coefficient of thermal expansion respectively. 
Throughout, open squares represent Lennard Jones results and open circles Morse results.}  
\end{figure}

The peak locations calculated in the framework of the analytical approximations (solid traces) for $c_{p}$ and $\alpha$ are in close agreement
with the exact numerical results (open symbols) as may be seen in the upper panels of Fig.~\ref{fig:Fig5}. 
In statistical mechanics singularities are not in general encountered for single component systems, with non-analytic
behavior emerging only in thermodynamic limit as the number of degrees of freedom tends to infinity.  Nevertheless, the 
low pressure regime is atypical in the sense that singular behavior is inevitable as $p$ (or $\gamma$) tends to zero 
due to the divergence of $\langle l \rangle$ for any finite $T$ as 
$p \rightarrow 0$. Hence, the possibility of a sharp dissociation transition for $\gamma \ll 1$ must be examined with care.

As a measure of the extent to which the dissociation transition is singular, the sharpness of 
the thermal expansion coefficient and specific heat peaks is quantified in the lower panels of 
Fig.~\ref{fig:Fig5} as the relative full width half maximum, $\Delta T_{\mathrm{Peak}}/T_{\mathrm{Peak}}$.
In the case of $\alpha$, $\Delta T_{\mathrm{Peak}}/T_{\mathrm{Peak}}$ tends to a finite value 
common to both $V_{\mathrm{LJ}}$ and $V_{\mathrm{M}}$, indicating the $\alpha$ peaks 
cease to become narrower relative to their location with decreasing $\gamma$.  

On the other hand, the specific heat relative peak width appears to tend to zero
with decreasing $\gamma$, a trend highlighted in the lower right panel inset of Fig.\ref{fig:Fig5}
showing $\Delta T_{\mathrm{Peak}}/T_{\mathrm{Peak}}$  relative to $1/\log_{10} (\gamma^{-1})$,
The curves are asymptotically linear as $\gamma \rightarrow 0$, with the relative 
peak width vanishing for $p \rightarrow 0$ as singular behavior appears.  

In conclusion, with a nonperturbative treatment of the anharmonicity of interatomic potentials, we provide 
a theoretical description of the dissociation transition valid for the condensed state as well as the gas-like phase where 
thermal fluctuations have driven pairs of atoms far from their equilibrium separations.
By applying a unified treatment to disparate potentials, we have obtained analytical 
results for salient thermodynamic observables in good agreement with precise numerical results.  Though 
ascribed to distinct bonding physics, there are striking similarities in results, e.g. with 
$\langle l \rangle/R_{e} = 1 + \varepsilon^{1/2} \exp -\varepsilon /(d_{1} \sqrt{\pi} \eta^{2})$  
specifying the mean bond length for $\gamma \ll 1$.

\begin{acknowledgments}
Useful conversations with Michael Crescimanno are gratefully acknowledged.
\end{acknowledgments}


\end{document}